\documentclass[a4paper,11pt]{article}
\usepackage{jcappub}
\usepackage{amsmath}

\makeatletter
\gdef\@fpheader{}
\makeatother

\newcommand{\sss}[1]{{\scriptscriptstyle{#1}}}

\newcommand{\deriv}[2]{#1_{\negthinspace,#2}}

\newcommand{\order}[1]{\mathcal{O}\!\left(#1\right)}

\newcommand{\dd}{\mathrm{d}}
\newcommand{\ud}{\dd}

\newcommand{\ie}{\textsl{i.e.}}

\newcommand{\urad}{\mathrm{rad}}

\newcommand{\urel}{\mathrm{rel}}
\newcommand{\ureh}{\mathrm{reh}}
\newcommand{\uend}{\mathrm{end}}
\newcommand{\uinf}{\mathrm{inf}}

\newcommand{\calB}{\mathcal{B}}

\newcommand{\MeV}{\mbox{MeV}}
\newcommand{\GeV}{\mbox{GeV}}

\newcommand{\Mpc}{\mbox{Mpc}}
\newcommand{\G}{\mathrm{G}}

\newcommand{\Mpl}{M_{_{\mathrm Pl}}}

\newcommand{\Prel}{P_\urel}

\newcommand{\Rrad}{R_\urad}

\newcommand{\Ereh}{E_\ureh}
\newcommand{\rhoreh}{\rho_\ureh}
\newcommand{\rhoend}{\rho_\uend}
\newcommand{\rhotildegamma}{\tilde{\rho}_{\gamma_0}}
\newcommand{\rhogamma}{\rho_{\gamma_0}}

\newcommand{\rhoB}{\rho_{\sss{B}}}
\newcommand{\rhoBzero}{\rho_{\sss{B_0}}}
\newcommand{\rhoBend}{\rho_{\sss{B_\uend}}}
\newcommand{\areh}{a_\ureh}

\newcommand{\aend}{a_\uend}
\newcommand{\zend}{z_\uend}
\newcommand{\Bzero}{B_0}

\newcommand{\Nend}{N_\uend}
\newcommand{\Nreh}{N_\ureh}

\newcommand{\Nstar}{N_*}
\newcommand{\Nzero}{N_0}
\newcommand{\wreh}{w_\ureh}
\newcommand{\Vend}{V_\uend}
\newcommand{\Vstar}{V_*}
\newcommand{\epsone}{{\epsilon_1}}

\newcommand{\epsend}{{\epsilon_1}_\uend}

\newcommand{\rdof}{\mathcal{Q}}
\newcommand{\gzero}{g_\sss{0}}
\newcommand{\greh}{g_\ureh}
\newcommand{\gs}{q}
\newcommand{\gszero}{\gs_\sss{0}}
\newcommand{\gsreh}{\gs_\ureh}

\newcommand{\zetainf}{\zeta^{(\uinf)}}
\newcommand{\zetareh}{\zeta^{(\ureh)}}
\newcommand{\zetarad}{\zeta^{(\urad)}}

\newcommand{\Pstar}{P_*}
\newcommand{\OmegaR}{\Omega_\gamma^0}

\title{Magneto-reheating constraints from curvature perturbations}

\author[a]{Christophe Ringeval,}
\author[b]{Teruaki Suyama}
\author[b,c]{and Jun'ichi Yokoyama}

\affiliation[a]{Centre for Cosmology, Particle Physics and Phenomenology,
  Institute of Mathematics and Physics, Louvain University, 2 Chemin
  du Cyclotron, 1348 Louvain-la-Neuve (Belgium)}

\affiliation[b]{Research Center for the Early Universe (RESCEU), Graduate School
  of Science,\\ The University of Tokyo, Tokyo 113-0033, Japan}

\affiliation[c]{Kavli Institute for the Physics and Mathematics of the
  Universe (Kavli IPMU),\\
  The University of Tokyo, Kashiwa, Chiba, 277-8568, Japan}

\emailAdd{christophe.ringeval@uclouvain.be}  
\emailAdd{suyama@resceu.s.u-tokyo.ac.jp}
\emailAdd{yokoyama@resceu.s.u-tokyo.ac.jp}

\abstract{As additional perturbative degrees of freedom, it is known
  that magnetic fields of inflationary origin can source curvature
  perturbations on super-Hubble scales. By requiring the magnetic
  generated curvature to remain smaller than its inflationary
  adiabatic counterpart during inflation and reheating, we derive new
  constraints on the maximal field value today, the reheating energy
  scale and its equation of state parameter. These bounds end up being
  stronger by a few order of magnitude than those associated with a
  possible backreaction of the magnetic field onto the background. Our
  results are readily applicable to any slow-roll single field
  inflationary models and any magnetic field having its energy density
  scaling as $a^\gamma$ during inflation. As an illustrative example,
  massive inflation is found to remain compatible with a magnetic
  field today $\Bzero = 5 \times 10^{-15}\,\G$ for some values of
  $\gamma$ only if a matter dominated reheating takes place at
  energies larger than $10^{5}\,\GeV$. Conversely, assuming
  $\gamma=-1$, massive inflation followed by a matter dominated
  reheating cannot explain large scale magnetic fields larger than
  $10^{-20}\,G$ today.}


\begin{document}

\maketitle

\section{Introduction}
\label{sec:intro}

Current measurements of magnetic fields filling the intergalactic
medium show that they are not vanishing on the largest length
scales. Using the spectra of blazars, and under the safest
assumptions, Ref.~\cite{Essey:2010nd} finds the two-sigma limit
$\Bzero > 10^{-17}\,\G$ whereas other data yield that $\Bzero >5\times
10^{-15}\,\G$~\cite{Dolag:2010ni, Neronov:1900zz, Tavecchio:2010mk}
and $\Bzero > 10^{-20}\,\G$~\cite{Inoue:2011zz}.  A non-vanishing
magnetic field today, and on the largest length scales, suggests that
it has a primordial origin. Indeed, as any other perturbation modes,
large scale magnetic fields are necessarily of super-Hubble
wavelengths at higher redshifts, and this triggers the problem of an
astrophysical generation
mechanism~\cite{Grasso:2000wj,Giovannini:2003yn, Kandus:2010nw}. Since
inflation solves exactly the same problem for the generation of the
primordial curvature perturbations, it sounds the best candidate to
generate magnetic field~\cite{Turner:1987bw}. However, in flat
space~\cite{Barrow:2011ic}, the conformal invariance of
electromagnetism prevents an amplification mechanism to take place
during inflation. Even worse, as the magnetic energy density redshifts
as $\rhoB \propto 1/a^4$, its contribution during inflation can become
dominant and leads to a severe backreaction
problem~\cite{Demozzi:2009fu, Kanno:2009ei, Byrnes:2011aa,
  Finelli:2011cw, Urban:2011bu}. In fact, backreaction can already
appear during the reheating era as soon as the mean equation of state
parameter $\wreh<1/3$. As shown in Ref.~\cite{Demozzi:2012wh}, this
yields some non-trivial constraints between the energy density
$\rhoreh$ at the end of the reheating era, the equation of state
parameter $\wreh$ and the present Hubble scale value of the magnetic
field $\Bzero$. As a result, if magnetic fields have an inflationary
origin, conformal invariance has to be broken, at least during
inflation~\cite{Ratra:1991bn, Dolgov:1993mu, Dolgov:1993vg,
  Tornkvist:2000js, Caprini:2001nb, Bamba:2003av, Bamba:2004cu,
  Ashoorioon:2004rs, DiazGil:2007dy, Martin:2007ue, Durrer:2010mq}.

Assuming the background evolution is under control, one still has to
consider the gravitational effects of the magnetic degrees of freedom
onto the evolution of the cosmological perturbations. Among others,
one expects the generation of primordial
non-Gaussianities~\cite{Barnaby:2012tk, Bartolo:2012sd,
  Shiraishi:2012rm, Shiraishi:2012xt, Shiraishi:2013vja} and a
non-conservation of the curvature perturbation on super-Hubble scales,
in a way similar to the presence of entropy modes~\cite{Bonvin:2011dt,
  Barnaby:2012tk, Suyama:2012wh, Giovannini:2013rme}. In this context,
magnetic fields should not induce too large curvature perturbations on
super-Hubble scales which would otherwise spoil the standard adiabatic
contribution of inflationary origin. In this paper, we look into this
issue, both during inflation and reheating, and use a phenomenological
model to describe the evolution of the magnetic field during inflation
as $B \propto a^\gamma$. Although our approach does not allow to
derive the backreaction of the perturbations onto the magnetic field
itself, our results should give the correct order of magnitude. In
particular, and in addition to $\gamma$, we show that the maximal
allowed value of $\Bzero$ on Hubble scales today depends on the way
the reheating proceeded and on the inflation model itself. By assuming
only slow-roll during inflation and a constant equation of state
during reheating, we derive a generic formulae giving the upper bound
of $\Bzero$ in terms of the energy scale of reheating $\rhoreh^{1/4}$
and the corresponding equation of state parameter $\wreh$. Such a
result therefore generalizes the background constraints derived in
Ref.~\cite{Demozzi:2012wh} and extend the results of
Ref.~\cite{Suyama:2012wh} to any reheating history. Our findings are
then applied to large field massive inflation and the small field
inflationary models. For instance, we find that massive inflation can
only be compatible with a matter dominated reheating and a magnetic
field value today of $\Bzero = 5\times 10^{-15}\,\G$ if the energy
scale of reheating is higher than $\Ereh \gtrsim 10^5\,\GeV$,
independently of $\gamma$. If some assumptions are made on $\gamma$,
the bounds can be much stronger and are represented in
Fig.~\ref{fig:lfp2}. Our main formulae are Eq.~(\ref{eq:zeta1inf}) and
Eqs.~(\ref{eq:zeta2inf}) to (\ref{eq:zeta2rad}) which explicitly give
the amplitude of the curvature perturbation given an inflationary
potential, the magnetic field today and the reheating parameter.

\section{Super-Hubble curvature perturbations from magnetic fields}

The super-Hubble scale curvature perturbation $\zeta$ on the constant
energy density hypersurface sourced by an inhomogeneous primordial
magnetic field is given by~\cite{Suyama:2012wh},
\begin{equation}
\zeta (t)=-\int_{t_*}^t dt_1 \frac{H(t_1)}{\rho(t_1)+P(t_1)} \delta
P_{\rm rel}(t_1)+\frac{8\pi G}{3} \int_{t_*}^t \frac{dt_1}{a^3 (t_1)}
\int_{t_*}^{t_1} dt_2~a^3(t_2) \Pi (t_2),
\label{eq:zeta-mag}
\end{equation}
where we have imposed an initial condition that $\zeta (t_*)=0$. In
the actual situation, $t_*$ may be taken to be a Hubble crossing time.
The first term, which we denote by $\zeta_1$, represents the
contribution due to the relative entropy perturbation, $\delta P_{\rm
  rel} \equiv \delta P_B-\frac{\dot P}{\dot \rho} \delta \rho_B$ where
$\delta P_B$ and $\delta \rho_B$ denote pressure and energy density
perturbation of the electromagnetic field. The quantities $P$ and
$\rho$ are the total pressure and the total energy density,
respectively. Assuming that the electromagnetic field obeys the
Maxwell equations after the end of inflation, $\zeta_1$ becomes
independent of time after reheating.  On the other hand, the second
term, which we denote by $\zeta_2$, is sourced by the anisotropic
stress of the magnetic field. Since the anisotropic stress persists
even after reheating, $\zeta_2$ still continues to evolve during and
after reheating.

\subsection{Phenomenological model}

In order to evaluate Eq.~(\ref{eq:zeta-mag}), we need to specify the time
evolution of $\delta \Prel$ and $\Pi$ during inflation, which
requires specification of the generation model of the magnetic field.
Since we do not want to concentrate on the particular model of the
magnetogenesis, in this paper, we will take a phenomenological
approach and make the following ansatz:
\begin{equation}
  \delta \Prel(t) = \alpha \rhoBend \left[\dfrac{a(t)}{\aend} \right]^\gamma,\qquad
\Pi(t) = \beta  \rhoBend \left[ \dfrac{a(t)}{\aend} \right]^\gamma,
\end{equation}
where $\alpha$, $\beta$ and $\gamma$ are constant parameters and
$\rhoBend$ is the energy density of the magnetic field at the end of
inflation, which is treated as first order perturbation.

Then, adopting the slow-roll approximation and assuming a constant
equation of state during reheating, Eq.~(\ref{eq:zeta-mag}) can be
explicitly integrated in terms of the number of e-folds $N$:
\begin{equation}
\begin{aligned}
  \zeta_1 = - \alpha \rhoBend \int_{\Nstar}^{\Nend}
  \dfrac{e^{\gamma(N-\Nend)}}{\dot{\phi}^2} \, \ud N -
  \dfrac{1}{3} 
   \frac{\rhoBend}{\rho_{\rm end}} \int_{\Nend}^{\Nreh} \frac{(1- 3\wreh)}{1+\wreh} \, e^{(-1+3\wreh)(N-\Nend)} \, \ud N,
\end{aligned}
\end{equation}
where a dot stands for differentiation with respect to $N$. The first
term can be further simplified by assuming slow-roll while the second
can be integrated explicitly. In reduced Planck mass units 
($\Mpl^2 =8 \pi G =1$), and at first order in the Hubble flow functions, one gets
\begin{equation}
\begin{aligned}
\zeta_1 & = 3 \alpha \rhoBend
\int_{\phi_*}^{\phi_\uend} 
\dfrac{V^2(\phi)}{\deriv{V}{\phi}^3(\phi)} \left[1 -
  \dfrac{\epsone(\phi)}{3}\right]  \exp{ \left[ \gamma
    \int_\phi^{\phi_\uend} \dfrac{V(\varphi)}{\deriv{V}{\varphi}(\varphi)} \ud
    \varphi \right]} \ud \phi + \dfrac{\rhoBend}{\rhoend}
\dfrac{\Rrad^4-1}{3(1+\wreh)} \,.
\end{aligned}
\label{eq:zeta1}
\end{equation}
In this expression
\begin{equation}
\label{eq:rrad}
\Rrad \equiv
\dfrac{\aend}{\areh}\left(\dfrac{\rhoend}{\rhoreh}\right)^{1/4} =
\left( \dfrac{\rhoreh}{\rhoend} \right)^{\frac{1-3\wreh}{12(1+\wreh)} },
\end{equation}
is the reheating parameter~\cite{Martin:2006rs, Ringeval:2007am,
  Martin:2010hh, Martin:2010kz}. The labels ``end'' and ``reh'' denote
respectively the end of inflation and the end of reheating, {\ie} the
beginning of the radiation era. Since after inflation we assume the
magnetic field to obey Maxwell equations, one can express the ratio
$\rhoBend/\rhoend$ in terms of energy densities
today~\cite{Demozzi:2012wh}.  In particular, one has $\rhoBend
=\rhoBzero (1+\zend)^4$ where $\zend$ is the redshift at the end of
inflation. Assuming instantaneous
transitions~\cite{Kahniashvili:2012vt} and making use of the reheating
parameter, one has
\begin{equation}
\label{eq:zend}
  1+\zend = \dfrac{1}{\Rrad} \left(\dfrac{\rhoend}{\rhotildegamma}
  \right)^{1/4},~~~~~\rhotildegamma \equiv \rdof_\ureh \rhogamma.
\end{equation}
Here $\rhogamma=3H_0^2 \OmegaR$ is the total radiation density today,
and $\rdof_\ureh \equiv \gszero^{4/3} \greh/(\gsreh^{4/3} \gzero)$ is the measure
of the change of relativistic degrees of freedom between the reheating epoch and today,
where $\gs$ and $g$ respectively denotes the number of entropy and
energetic relativistic degrees of freedom at the epoch of
interest. Plugging Eq.~(\ref{eq:zend}) into Eq.~(\ref{eq:zeta1}),
and making use of the Friedmann--Lema\^{\i}tre equations to express
$\rhoend$ in terms of the field potential $\Vend \equiv V(\phi_\uend)$,
one finally gets
\begin{equation}
\label{eq:zeta1inf}
\begin{aligned}
\zeta_1 = \frac{1}{\Rrad^4} \dfrac{\rhoBzero}{\rhotildegamma}
\left[\dfrac{9 \alpha}{3 - \epsend}
  \int_{\phi_*}^{\phi_\uend}  \dfrac{\Vend
    V^2}{\deriv{V}{\phi}^3} \left(1 - \dfrac{\epsone}{3} \right)
  e^{\gamma \Delta N(\phi)} \ud \phi + \dfrac{\Rrad^4 - 1}{3(1+\wreh)} \right].
\end{aligned}
\end{equation}
Here $\Delta N(\phi)<0$ is the number of e-folds before the end of
inflation and can be obtained from the slow-roll trajectory
\begin{equation}
\label{eq:srtraj}
  \Delta N(\phi) \equiv N(\phi) - \Nend \simeq -\int_{\phi_\uend}^{\phi}
  \dfrac{V}{\deriv{V}{\varphi}} \ud \varphi\,.
\end{equation}
The first Hubble flow function $\epsone(\phi)$ is evaluated along the
field trajectory and is also uniquely determined by the potential in
the slow-roll approximation~\cite{Schwarz:2001vv}
\begin{equation}
  \epsone(\phi) = \dfrac{\dot{\phi}^2}{2H^2} \simeq \dfrac{1}{2}
  \left(\dfrac{\deriv{V}{\phi}}{V}\right)^2.
\end{equation}
By definition, the quantity $\epsend \equiv \epsone(\phi_\uend)$ is
unity for inflationary models ending by slow-roll violation but can be
much smaller than unity for inflationary models ending by tachyonic
instabilities. This formula allows us to evaluate $\zeta_1$ once the
inflation model and the reheating are specified.

In a similar manner, straightforward calculations yield $\zeta_2$ at
any e-fold $N$ during the radiation era. For this, it is convenient to
split the integral over the three domains, inflation for
$\Nstar<N<\Nend$, reheating for $\Nend< N < \Nreh$ and radiation era
with $N>\Nreh$:
\begin{equation}
\zeta_2(N) = \zetainf_2 + \zetareh_2 + \zetarad_2(N).
\end{equation}
Again assuming slow-roll during inflation, and a constant
equation of state during reheating, one gets, at first order in the
Hubble flow functions:
\begin{align}
\label{eq:zeta2inf}
  \zetainf_2 & = - \dfrac{1}{\Rrad^4}
  \dfrac{\rhoBzero}{\rhotildegamma} \dfrac{3 \beta}{(3+\gamma)(3 -
    \epsend)} \int_{\phi_*}^{\phi_\uend}
  \dfrac{\Vend}{\deriv{V}{\phi}} \Bigg[ \left(1 -
    \dfrac{6+\gamma}{9+3\gamma} \epsone \right) e^{\gamma \Delta
    N(\phi)} \nonumber \\
  & - \sqrt{\dfrac{V}{V_*}} \left(1 - \dfrac{1}{6} \epsone \right)
  e^{-3 \Delta N(\phi)} e^{(3+\gamma) \Delta\Nstar} \Bigg] \ud \phi\,,
  \\
\label{eq:zeta2reh}
  \zetareh_2 & = \dfrac{2 \beta}{\Rrad^4}
  \dfrac{\rhoBzero}{\rhotildegamma} \Bigg\{ - \dfrac{\Rrad^4 -1}{1-9
    \wreh^2} + \dfrac{\left(\Rrad^2\right)^{\frac{3\wreh-3}{3 \wreh
        -1}} -
    1}{(3+\gamma)(3 \wreh-3)} \nonumber \\
  & \times \left[1 - \dfrac{\epsend}{3+\gamma} - \dfrac{6+2 \gamma}{3
      \wreh+1} -\sqrt{\dfrac{3\Vend}{(3-\epsend)\Vstar}} \,
    e^{(3+\gamma)\Delta
      \Nstar} \right] \Bigg\}\,,\\
\label{eq:zeta2rad}
  \zetarad_2(N) & = \dfrac{\beta}{\Rrad^4}
  \dfrac{\rhoBzero}{\rhotildegamma} \Bigg\{
  \Rrad^4(N - \Nreh) + \dfrac{1-3\wreh}{1+3\wreh} \Rrad^4   \nonumber \\
  & + \dfrac{(\Rrad^2)^{\frac{3 \wreh-3}{3 \wreh-1}}}{3+\gamma}
  \left[1 - \dfrac{\epsend}{3+\gamma} - \dfrac{6+2 \gamma}{3 \wreh+1}
    - \sqrt{\dfrac{3\Vend}{(3-\epsend)\Vstar}} \, e^{(3+\gamma)\Delta
      \Nstar} \right] \Bigg\}\,.
\end{align}
As before, $\epsone(\phi)$ stands for the first Hubble flow function
evaluated along the field trajectory and
$\epsend=\epsone(\phi_\uend)$. The quantity $\Delta\Nstar \equiv
\Nstar - \Nend <0$ is the number of e-folds before the end of
inflation at which the pivot scale $k_*$ crossed the Hubble radius
during inflation. It does not depend on $\phi$ in the previous
equations but, as discussed below, it is an inflationary
model-dependent function of the reheating parameter $\Rrad$ (see
Ref.~\cite{Martin:2010kz}). The same remark holds for $V_* \equiv
V(\phi_*)$. Finally, when relevant, we have dropped all terms
involving ${\epsone}_*$ as we always have ${\epsone}_* \ll 1$ for all
inflationary models.

Let us stress that the quantity $\zetarad_2(N)$ grows during the
radiation era, as $N-\Nreh$. However, Eq.~(\ref{eq:zeta2rad}) only
makes sense if the perturbation mode under consideration is
super-Hubble. As a result, an upper bound for $N-\Nreh$ is given by the
number of e-fold \emph{after} the end of reheating at which the pivot
scale $k_*$ re-enters the Hubble radius.


Finally, let us notice that the expressions for $\zeta_1$,
$\zetainf_2$, $\zetareh_2$ and $\zetarad_2$ are all proportional to
the factor $\rhoBend/\rhoend=\Rrad^{-4}\rhoBzero/\rhotildegamma$. As
shown in Ref.~\cite{Demozzi:2012wh}, imposing this factor to be
smaller than unity is equivalent to avoid magnetic field backreaction
over the background energy density during reheating. As here we are
requiring that $|\zeta|<10^{-5}$, our constraints are expected to be
typically, and at least, $2.5$ order of magnitude stronger than those
derived from the background evolution (see Fig.~\ref{fig:lfp2}).

\subsection{Reheating consistent slow-roll}

As already mentioned, both $\zeta_1$ and $\zeta_2$ depends on
parameters of the inflationary model as well as the reheating energy
scale. In particular, one needs to determine the value of all ``$*$''
quantities, {\ie} evaluated $\Delta\Nstar$ e-folds before the end of
inflation. These parameters cannot be freely chosen as one wants them
to be compatible with the observed amplitude of the CMB
fluctuations. In fact, as shown in Ref.~\cite{Martin:2006rs},
$\Delta\Nstar$ is itself a function of $\Rrad$. In order to ensure
consistency with reheating, we here follow the slow-roll approach of
Ref.~\cite{Martin:2010kz} that we briefly recap.

The e-fold $\Nstar$ is by definition solution of $k_*/a(\Nstar)
=H(\Nstar)$ during inflation. This equation can be recast in terms of
``observable'' quantities, namely
\begin{equation}
\label{eq:nstar}
\dfrac{k_*}{a_0} (1+\zend) e^{-\Delta\Nstar} = H_*\,.
\end{equation}
The right hand side of this equation can be fixed by the amplitude of
the adiabatic primordial power spectrum, which is a well measured
quantity. At leading order in slow-roll
\begin{equation}
\Pstar = \dfrac{H_*^2}{8 \pi^2 \epsone_* \Mpl^2}\,.
\end{equation}

From Eqs.~(\ref{eq:rrad}) and (\ref{eq:zend}), $\zend$ can be
expressed in terms of $\Rrad$ and $\rhoend$, the energy density at the
end of inflation. The latter can, in turn, be further simplified using
the Friedmann--Lema\^{\i}tre equations for a scalar field (in Planck
units)
\begin{equation}
\label{eq:rhoend}
\rhoend = \dfrac{3\Vend}{3-\epsend} = \dfrac{\Vend}{V_*}\dfrac{3
  V_*}{3-\epsend} = 3 H_*^2 \dfrac{\Vend}{V_*} \dfrac{3-\epsone_*}{3-\epsend}\,.
\end{equation}
As before, one can drop the term in $\epsone_* \ll 3$. The advantage
of this last expression is that it does no longer depend on the
potential normalization but involves only $H_*$.
Plugging everything back into Eq.~(\ref{eq:nstar}), 
$\Delta\Nstar$ is a solution of the algebraic equation~\cite{Martin:2006rs}
\begin{equation}
\Delta\Nstar = -\ln \Rrad + \Nzero + \dfrac{1}{4}
\ln\left[\dfrac{9}{\epsone_*(3-\epsend)} \dfrac{\Vend}{V_*}\right] -
\dfrac{1}{4} \ln (8 \pi^2 \Pstar)\,. \label{sol-alg}
\end{equation}
Here the quantity $|\Nzero|$ roughly measures the number of e-folds of
deceleration and is defined in Planck units by
\begin{equation}
\Nzero \equiv \ln \left[\dfrac{k_*/a_0}{\left(3 \rdof_\ureh \OmegaR
      H_0^2 \right)^{1/4}}\right].
\end{equation}
Notice that the trajectory is needed to evaluate the right hand side of Eq.~(\ref{sol-alg})
as $V_*$ and $\epsone_*$ are functions of $\phi(\Nstar)$. One can
nevertheless render this equation more explicit in terms of $\phi_*$
and $\rhoreh$ by assuming slow-roll and expanding $\Rrad$ from
Eqs.~(\ref{eq:rrad}) and (\ref{eq:rhoend}), {\ie}
\begin{equation}
\label{eq:lnRrad}
\ln \Rrad = \dfrac{1-3\wreh}{3+3\wreh} \ln \left(\rhoreh^{1/4} \right)
- \dfrac{1-3\wreh}{12(1+\wreh)} \ln \left[\dfrac{9
    \epsone_*}{3-\epsend} \dfrac{\Vend}{V_*} \right] -
\dfrac{1-3\wreh}{12(1+\wreh)} \ln\left(8 \pi^2 \Pstar\right).
\end{equation}
One finally gets
\begin{equation}
\label{eq:phistar}
\begin{aligned}
  \Delta \Nstar = \int_{\phi_*}^{\phi_\uend}
  \dfrac{V}{\deriv{V}{\varphi}} \ud \varphi & =
  -\dfrac{1-3\wreh}{3+3\wreh} \ln(\rhoreh^{1/4}) + \Nzero -
  \dfrac{1+3\wreh}{2(3+3\wreh)} \ln(8\pi^2 \Pstar) \\ & + \dfrac{1}{3
    + 3\wreh} \ln\left[\dfrac{9}{(\epsone_*)^{\frac{3\wreh+1}{2}}}
    \dfrac{\Vend}{(3 - \epsend) V_*}\right].
\end{aligned}
\end{equation}
Once the slow-roll trajectory is integrated, this expression is explicit
in $\phi_*$ and can be solved by specifying only $\Ereh \equiv \rhoreh^{1/4}$ and $\wreh$. 
Plugging the result into
Eqs.~(\ref{eq:zeta2inf}), (\ref{eq:zeta2reh}), (\ref{eq:zeta2rad}),
(\ref{eq:lnRrad}), (\ref{eq:phistar}) and finally Eq.~(\ref{eq:zeta1inf})
allows to determine $\zeta(N_*)/\rhoBzero$ uniquely from the input of
$\Ereh$ and $\wreh$. Imposing that $|\zeta(N_*)|^2 \ll \Pstar$
actually yields the reheating-dependent upper bound on $\rhoBzero$.

\section{Application to some representative models}
\subsection{Large field models}
\label{sec:lf}
The potential energy for the large field models is given by
\begin{equation}
\label{eq:potlf}
V(\phi)=M^4 \phi^p,
\end{equation}
where $M$ is a constant of mass dimension (in Planck unit) and $p$ is a positive number.
In order to be definite, we assume that this potential is correct not only
for large $\phi$ responsible for inflation but also for small $\phi$ relevant
for oscillating period and reheating.
For this potential, the field value at which inflation terminates is given by
\begin{equation}
\label{eq:phiendlf}
\phi_{\uend}=\frac{p}{\sqrt{2}}\,,
\end{equation}
the solution of $\epsone(\phi_\uend) = 1$ where
\begin{equation}
\label{eq:eps1lf}
  \epsilon_1(\phi) \simeq \dfrac{p^2}{2 \phi^2}\,.
\end{equation}
The slow-roll evolution of $\phi$ during inflation is given by
integrating Eq.~(\ref{eq:srtraj}). In Planck units, one gets
\begin{equation}
\label{eq:trajlf}
\Delta N = \dfrac{1}{2p}\left(\phi_\uend^2 - \phi^2\right),
\end{equation}
where, as before $\Delta N=N-\Nend$, is the number of e-fold measured
from the end of inflation.  After inflation, $\phi$ oscillates around
the minimum such that the natural equation of state parameter is given
by $\wreh=(p-2)/(p+2)$~\cite{Turner:1983he}.

Plugging Eqs.~(\ref{eq:potlf}), (\ref{eq:phiendlf}), (\ref{eq:eps1lf})
and (\ref{eq:trajlf}) into Eq.~(\ref{eq:zeta1inf}) and
Eqs.~(\ref{eq:zeta2inf}) to (\ref{eq:zeta2rad}) gives the expression
of $\zeta(N_*)/\rhoBzero$ in terms of the potential parameter $p$,
$\Rrad$ and $\phi_*$ only (plus the magnetic parameters). What remains
to do is to solve Eq.~(\ref{eq:phistar}) to get $\phi_*$ in terms of
$p$ and $\rhoreh$, from which $\Rrad$ is determined by
Eq.~(\ref{eq:lnRrad}).

For the large field model, an analytic solution of
Eq.~(\ref{eq:phistar}) can be found in terms of the Lambert
function~\cite{Martin:2010kz}, but as the various integrations
entering into the expression of $\zeta(N_*)$ that can only be
performed numerically, we have here preferred to solve this equation
numerically. The various cosmological parameters have been set to
their preferred values from the WMAP data~\cite{Komatsu:2010fb,
  Hinshaw:2012fq, Bennett:2012fp}, {\ie} $\Pstar \simeq 2.16\times
10^{-9}$, $h \simeq 0.72$, $\OmegaR = 4.6\times 10^{-5}$, and we have
set all $\rdof$ to unity for simplicity (they have only a small
effect). With a pivot scale chosen at $k_* = 0.05\,\Mpc^{-1}$, one has
$\Nzero \simeq -62$. The final result is a parametric curve $\Bzero =
\calB_{\alpha,\beta,\gamma,p}(\Ereh,\wreh)$, solution of
\begin{equation}
\label{eq:zetamax}
\zeta(N_*) = \sqrt{P_*}\,,
\end{equation}
which separates the plane $(\Ereh,\Bzero)$ in two regions. Above this
curve, the combination of the magnetic field value today and the
energy scale of reheating would be such that the magnetically generated
super-Hubble curvature perturbation would be equal or larger than the
adiabatic modes generated during inflation. The allowed region
therefore lies under this curve. Let us notice that this is a very
conservative upper limit as current constraints on isocurvature modes show
that they cannot exceed $10\%$ of the adiabatic
counterparts~\cite{Trotta:2006ww, Sollom:2009vd}.

\begin{figure}
\begin{center}
\includegraphics[width=0.8\textwidth]{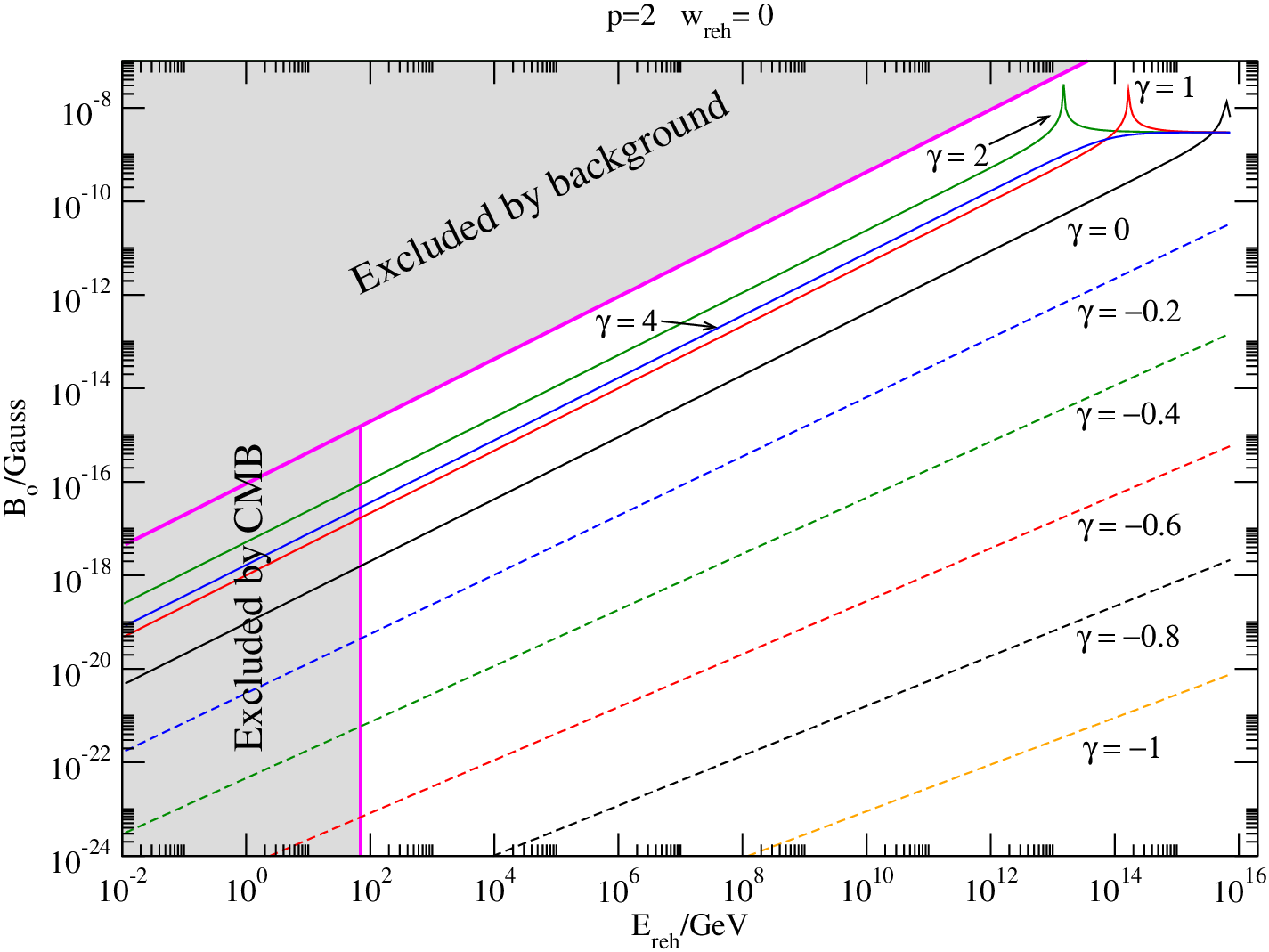}
\end{center}
\caption{Limits on the amplitude of the present magnetic field of
  inflationary origin with respect to the reheating energy scale. We
  have assumed a large field model with $p=2$ (and generic values for
  $\alpha=4/3$, $\beta=1$). Each line is the solution of
  Eq.~(\ref{eq:zetamax}) for various values of $\gamma$. The region
  above the corresponding line is excluded. The reheating energy scale
  $\Ereh\equiv \rhoreh^{1/4}$ varies from $100\,\MeV$ (BBN energy
  scale) to $\rhoend^{1/4}$, the energy at which inflation ends, see
  Eq.~(\ref{eq:rhoend}). For convenience, the CMB lower bound for the
  large field with $p=2$ reheating energy has been reported
  $\Ereh>70\,\GeV$ ($95\%)$, see Ref.~\cite{Martin:2010hh}. We have
  also represented the region excluded by magnetic field backreaction
  over the background energy density during reheating, see
  Ref.~\cite{Demozzi:2012wh}.}
 \label{fig:lfp2}
\end{figure}

In Fig.~\ref{fig:lfp2}, we have plotted these limits for the large
field model with $p=2$ and for various values of $\gamma$ (taking
$\alpha=4/3$ and $\beta=1$ as reference values of $\order{1}$
parameters).  Negative values of $\gamma$ provides the tightest bounds
on the magnetic field. This is expected since in that situation the
magnetic contribution increases more and more for $a \rightarrow 0$,
{\ie} deep during inflation. As it is already known, standard values
for $\gamma=-4$ during inflation would even generate a strong
backreaction problem~\cite{Demozzi:2009fu, Kanno:2009ei,
  Urban:2011bu}. Here, even for $\gamma=-1$, curvature perturbations
generated by the magnetic fields become important during inflation and
$\zeta(N_*) \simeq \zeta_1 +\zetainf_2$. We see on the figure that,
for $\gamma=-1$, one cannot actually generate a magnetic field today
larger than $\Bzero =10^{-20}\,\G$.  In the opposite situation,
$\gamma>0$, deep during inflation the magnetic effects are very small
and $\zeta(N_*) \simeq \zetareh_2 + \zetarad_2(N_*)$ is mostly
generated after inflation. Remembering that after inflation $\rhoB
\propto a^{-4}$, at fixed $\Bzero$, a longer period of reheating,
{\ie} a lower reheating temperature, results in higher magnetic field
values at the end of inflation; and hence a larger $\zeta$. As for the
background case discussed in Ref.~\cite{Demozzi:2012wh}, this effect
is enhanced if the energy density of the universe decreases more
slowly than radiation, as this is the case for $\wreh=0$ here. As can
be checked in Eqs.~(\ref{eq:zeta2reh}) and (\ref{eq:zeta2rad}), the
$\zeta$-dependency of these two terms with respect to $\gamma$ is very
weak (as opposed to $\zetainf_2$). In Fig.~\ref{fig:lfp2}, one indeed
sees that the bounds become almost insensitive to $\gamma$ as soon as
$\gamma\gtrsim 2$ such that this limit is actually very conservative
and somehow model-independent.  At around $\Ereh = 10^{14}\,\GeV$,
a spiky bump is observed for $\gamma=1$ curve (and similar one for
$\gamma=0$ curve at higher value of $\Ereh$.).  This is due to the
occasional vanishing of $\zeta$, which is possible since $\zeta$ is a
function of $\Rrad$ and hence of $\Ereh$.  At that point,
$\zeta$ vanishes irrespective of the amplitude of the magnetic field
and, as a result, we have no constraint on the magnetic field, which
shows up as the bump.

\begin{figure}
  \begin{center}
    \includegraphics[width=0.8\textwidth]{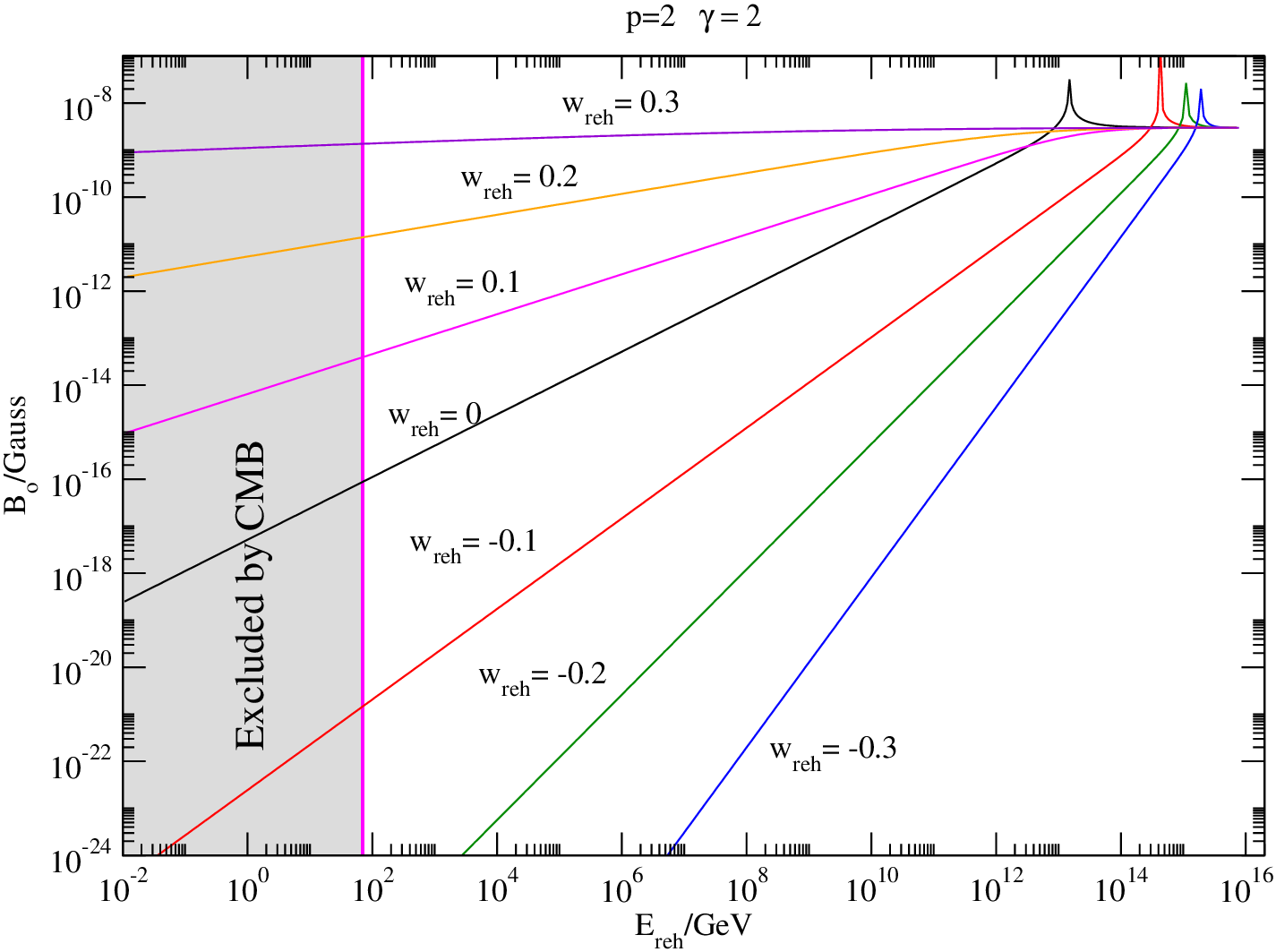}
  \end{center}  
    \caption{Limits on the amplitude of the present magnetic field of
      inflationary origin with respect to the reheating energy scale for a
      large field model $p=2$, $\gamma=2$ and for various values of
      $\wreh$.}
  \label{fig:lfp2_wreh}
\end{figure}

In Fig.~\ref{fig:lfp2_wreh}, we have plotted the magneto-reheating
constraints, still for a large field model with $p=2$, but assuming
another equation of state parameter $\wreh$. As expected, the more
$\wreh$ becomes negative, the tighter the limits are whereas there is
no constraints for a radiation-like reheating era.

\subsection{Small field models}

\begin{figure}
  \begin{center}
    \includegraphics[width=0.8\textwidth]{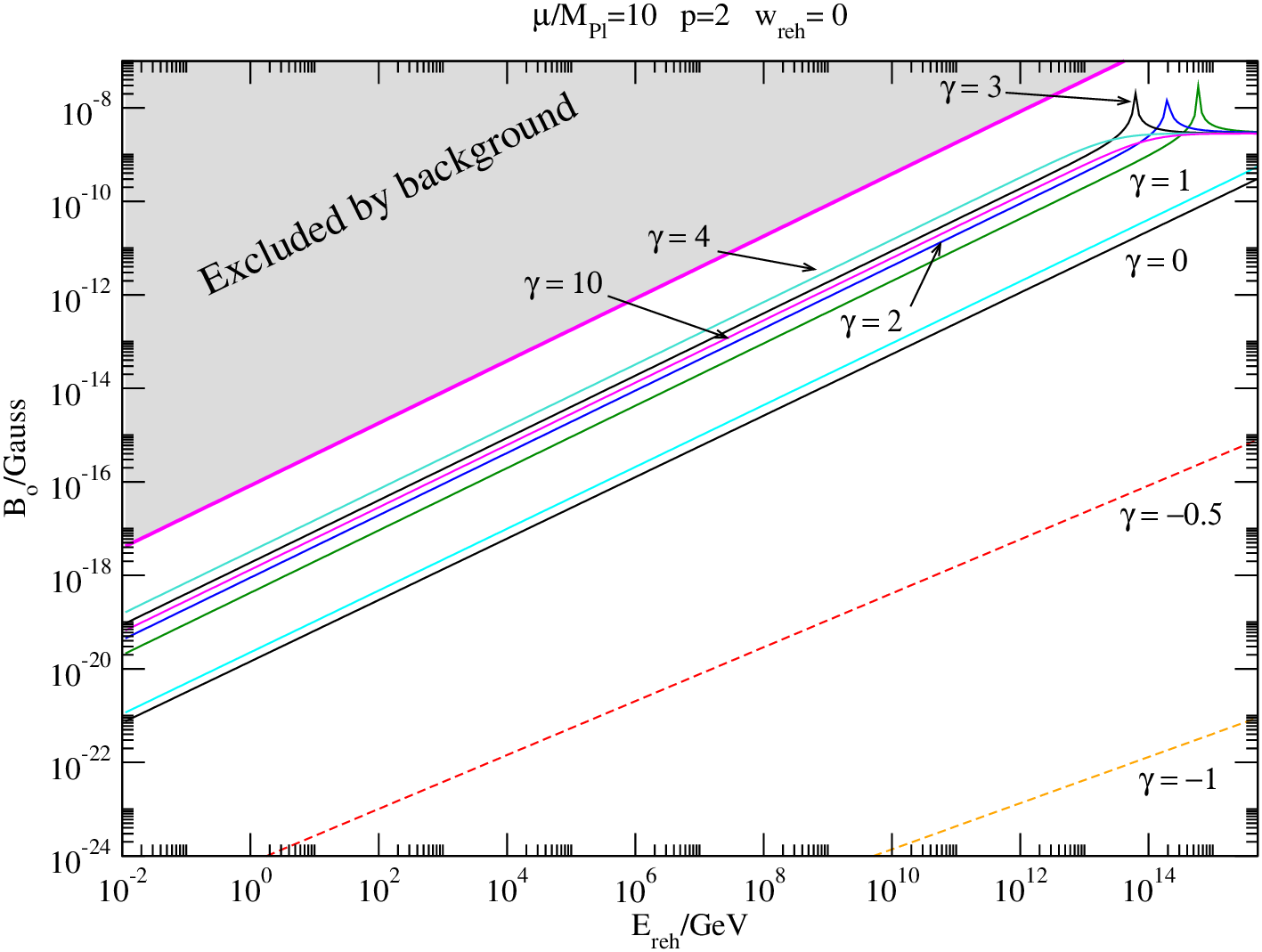}   
    \caption{Limits on the amplitude of the present magnetic field of
      inflationary origin assuming the small field models with $p=2$
      and $\mu=10\,\Mpl$. For each labelled value of $\gamma$, the
      region above the corresponding line is excluded. These bounds
      are almost unchanged for larger values of $\mu$ (see text).}
    \label{fig:sfp2}
  \end{center}
\end{figure}

Let us next consider the small field models for which the potential is
given by
\begin{equation}
V(\phi)=M^4 \left[ 1-{\left( \frac{\phi}{\mu} \right)}^p \right],
\end{equation}
where $M$ and $\mu$ are constants of mass dimension in Planck unit and $p$ is a
positive number. Notice that this potential is valid only for the
inflationary regime $\phi/\mu \ll 1$ and has to be replaced by another
function after the end of inflation. If this is a quadratic function,
one can assume $\wreh\simeq 0$. For this potential, normalizing the
field value by $\mu$ as $\chi= \phi/\mu$, the slow-roll trajectory
reads (in Planck units)
\begin{equation}
  \Delta N=-\frac{\mu^2}{2 p} \left[\left(\chi^2 +
      \dfrac{2}{2-p}\chi^{2-p}\right) - \left(
      \chi_\uend^2 + \dfrac{2}{2-p} \chi_\uend^{2-p} \right) \right],
    \label{eq:trajsf}
\end{equation}
where $\chi_\uend$ is determined by $\epsone(\chi_\uend)=1$. From
\begin{equation}
\label{eq:sfeps}
\epsone(\chi) = \dfrac{1}{2} \left(\dfrac{p}{\mu} \dfrac{\chi^{p-1}}{1
    - \chi^p} \right)^2,
\end{equation}
one finds a transcendental equation for $\chi_\uend$:
\begin{equation}
  \chi_\uend^{p-1} = \sqrt{2} \dfrac{\mu}{p} \left(1-\chi_\uend^p \right).
\end{equation}
Equation~(\ref{eq:trajsf}) can also be applied to the case $p=2$ by
taking the limit $p \to 2$, which yields terms in $\ln (\chi
/\chi_\uend)$. Along the lines detailed for the large field models,
these expressions are enough to completely determine the upper bound
$\Bzero=\calB(\Ereh,\wreh)$ parametrized by $\mu$ and $p$ (see
Sect.~\ref{sec:lf}).

\begin{figure}
  \begin{center}
    \includegraphics[width=0.8\textwidth]{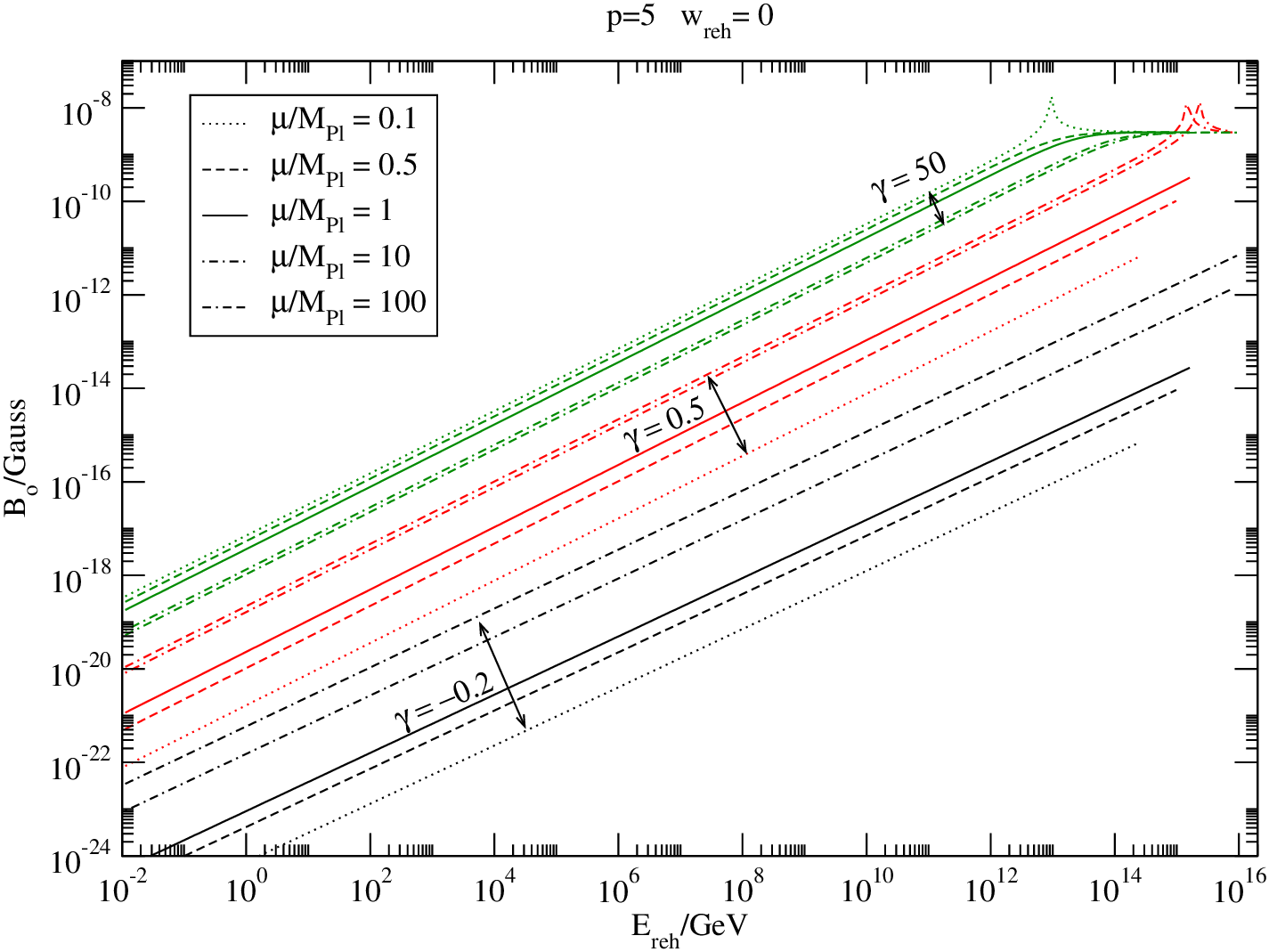}
    \includegraphics[width=0.8\textwidth]{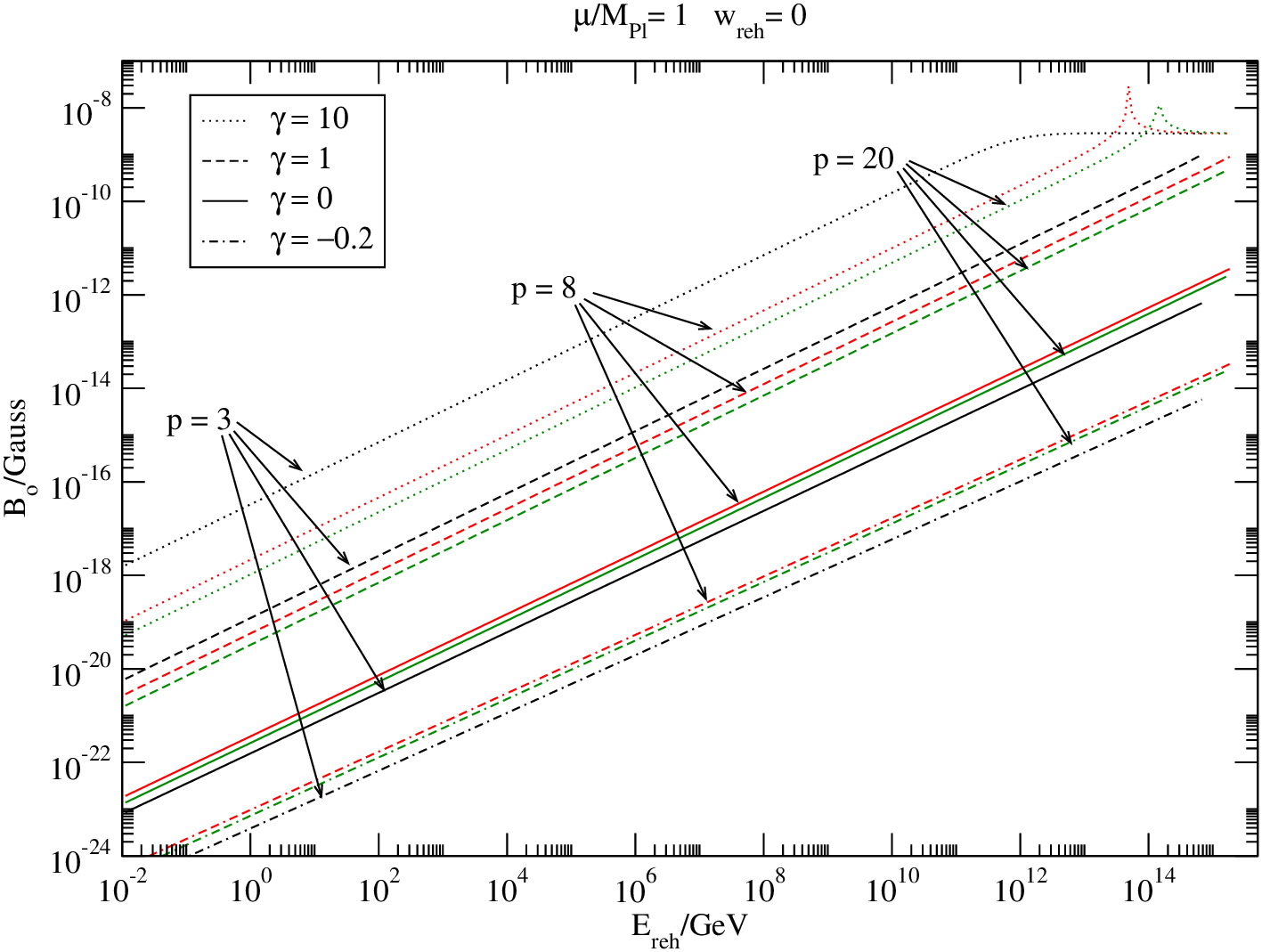}
    \caption{This panel shows the limit on the amplitude of the
      present magnetic field of inflationary origin assuming the small
      field models with with various values of $p$ and at fixed
      $\mu=\Mpl$. As before, the region above the lines is excluded and
      $\alpha=\frac{4}{3}$, $\beta=1$.}
    \label{fig:sfscan}
  \end{center}
\end{figure}

In Fig.~\ref{fig:sfp2}, we have plotted the limits on the amplitude of
the present magnetic field of inflationary origin as a function of the
reheating temperature assuming small field models with $p=2$ and
$\mu=10\Mpl$. Notice that for $\wreh=0$, there is currently no bounds
on the reheating energy scale coming from CMB
alone~\cite{Martin:2010hh}. Let us also recap that the models having
$p=2$ are compatible with the current spectral index and
tensor-to-scalar ratio only for large values of $\mu$, typically $\mu
\gg \Mpl$~\cite{Martin:2006rs}. The bounds of Fig.~\ref{fig:sfp2}
exhibits the same behaviour as the large field models, {\ie} tighter
bounds on the magnetic field for lower reheating temperatures and for
lower values of $\gamma$. We also find that those results remain
insensitive to larger values of $\mu$. This is not surprising since,
for the small field models, the first two-Hubble flow functions become
independent of $\mu$ in the large
$\mu$-limit~\cite{Martin:2010hh}. Moreover, the above equations imply
that $1-\chi_\uend=\order{\Mpl/\mu}$ and
$\chi_\uend-\chi_*=\order{\Mpl/\mu}$. Therefore the integral in
Eq.~(\ref{eq:zeta1inf}) scales as $\Mpl^3/\mu^3$ and hence $\zeta_1$
becomes also independent of $\mu$ in the large $\mu$-limit.

In the top panel of Fig.~\ref{fig:sfscan}, the limits on the amplitude
of the magnetic field have been derived for the case $p=5$, and we
have made both $\gamma$ and $\mu$ vary. Contrary to the case $p=2$,
sub-Planckian values for $\mu$ can be made compatible with CMB
data~\cite{Martin:2010hh}. The magneto-reheating bounds become
generically stronger at small $\mu$; but such a behaviour can be
transiently inverted for some very large values of $\gamma \gg 1$. For
instance, this is the case for $\gamma=50$, but for values of
$\mu<10^{-2}$ the upper bound moves again downwards in
Fig.~\ref{fig:sfscan} (not represented). As before, for $\gamma<0$ the
bounds are driven by the behaviour of $\zeta_1$ because the magnetic
effects are dominant deep in inflation. For $\gamma>0$, magnetic
effects during reheating are the most important, and $\zetareh_2$ is
the dominant term such that the $\mu$-dependence ends up being related
to the values of $\gamma$. In the limit $\mu/\Mpl \ll 1$, one can
nevertheless use some crude approximations to guess the dependency in
$\mu$. One has $\chi_* \simeq [p(p-2) \mu^2 / |\Delta
\Nstar|]^{1/(p-2)}$ such that the dominant terms in
Eq.~(\ref{eq:zeta1inf}) scale as (in Planck units)
\begin{equation}
\int_{\chi_*}^{\chi_\uend} \dfrac{\mu^4}{p^3} \chi^{3(1-p)}
\exp\left[-\gamma \dfrac{\mu^2}{p(p-2)} \chi^{2-p} \right] \ud \chi
\propto \mu^{-2p/(p-2)}\,.
\end{equation}
Similarly, one can use Eq.~(\ref{eq:lnRrad}) to get 
$\Rrad^4 \propto \epsone_*^{(3 \wreh-1)/(3+3\wreh)}$. In the limit $\mu\ll 1$,
Eq.~(\ref{eq:sfeps}) shows that $\epsone_* \propto \mu^{2p/(p-2)}$ and, 
for $\wreh=0$, one finally gets that $\zeta_1 \propto \epsone_*^{-2/3}$. 
As expected, this quantity increases as $\mu$ decreases. 
For $\gamma>0$, the behaviour of $\zeta_2$ is now
driven by powers of $\Rrad$, which again increases when $\mu$
decreases, albeit in a different way.

In the bottom panel of Fig.~\ref{fig:sfscan}, we have fixed $\mu=\Mpl$
and plotted the upper limits for $p=3$, $p=8$ and $p=20$. Since $p$
only appears in the equation as an order unity factor, the final
dependency in $p$ remains weak. Notice again the change of behaviour
between large and low values of $\gamma$ that can, as for $\mu$, be
traced back to which part of $\zeta$ contributes the most.

\section{Conclusion}
Recent observations of the cosmic rays reveal that the magnetic fields
are ubiquitous in the universe.  Lack of the convincing astrophysical
explanation for the origin of such magnetic fields has led some
theorists to seriously consider the possibility that those magnetic
fields are produced during the primordial inflation.  In addition to
the obvious condition on the model of the inflationary magnetogenesis
that it must produce the observed amplitude of the magnetic field at
the relevant scales, there is another condition that must be taken
into account for whatever the model of the magnetogenesis is.  Since
the inhomogeneous magnetic fields enter the right hand side of the
Einstein equation as perturbations of the energy-momentum tensor, they
induce the metric perturbation in addition to the standard adiabatic
one produced by the inflaton fluctuation.  The condition must be
satisfied that this additional metric perturbation should not exceed
the observed one in order for any inflationary magnetogenesis model to
be embedded consistently in the standard model of the early universe.

In this paper, we first derived general expression of the resulting
super-horizon scale curvature perturbation sourced by the primordial
magnetic field, without resorting to the specific model of
inflationary magnetogenesis but with a phenomenological assumption
that the magnetic field energy density and the associated anisotropic
stress scale as $a^\gamma$, where $\gamma$ is a free parameter, so
that our result provides a wide coverage for any magnetogenesis model
satisfying the above scaling behavior.  Our result can be also applied
to any canonical single field inflation model followed by the
oscillations of the inflaton with any equation of state and by
reheating with any reheating energy scale.  Given the inflation model
and the measurement of the curvature power spectrum at the pivot scale,
our formula allows us to put bound on the combination of the amplitude
of the today's magnetic field, reheating energy scale and its equation
of state parameter.  In practice, our formula requires numerical
solution of the algebraic equation to find the inflaton field value
corresponding to the Hubble crossing of the pivot scale and one
dimensional numerical integration for evaluating the curvature
perturbation, both of which are quite feasible to implement.  Our
perturbation bound is tighter by a few orders of magnitude than the
one given in the literature, which is derived from the requirement
that the magnetic field energy density be smaller than the background
energy density so that it does not destroy the homogeneous and
isotropic Friedmann--Lema\^{\i}tre metric.

We then applied our formula to the large field inflation models and
the small field inflation models, which are representative models of
inflation.  In either case, the upper bound on the value of the
magnetic field today becomes tighter for negative value of $\gamma$,
where the magnetic curvature perturbation is dominantly produced deep
inside inflation.  In particular, the magnetic field strength of
$10^{-15}\,\G$, which has the observational relevance, with
$\gamma < -1$ is completely incompatible with the perturbation bound.
On the other hand, the upper bound is insensitive for positive
$\gamma$ and saturates at $\gamma = \order{1}$.  Thus we can
interpret the upper bound for $\gamma=\order{1}$ as the fairly
conservative bound.  With $\gamma$ being fixed, the upper bound on the
magnetic field becomes severer as the reheating energy scale becomes
lower if the reheating equation of state parameter is less than $1/3$.
This is expected since the magnetic field energy density contributes
more to the total energy density as we go back into the past.  For
example, for the chaotic inflation model with a quadratic potential,
the magnetic field today must be less than $10^{17} \,\G$ if
the reheating energy scale is as low as $70\,\GeV$ that is the
possible minimal energy scale allowed by WMAP 7yr data.  If the cosmic
magnetic fields actually turn out to be of the inflationary origin in
the future and search for the realistic model of the inflationary
magnetogenesis becomes an indispensable pillar in constructing a
realistic evolutionary scenario of the early universe, our
perturbation bound derived in this paper provides one of the necessary
conditions that must be considered for constraining the magnetogenesis
models.

\acknowledgments

This work was supported by a Grant-in-Aid for JSPS Fellows
No.~1008477(TS), JSPS Grant-in-Aid for Scientific Research No.\
23340058 (JY), the Grant-in-Aid for Scientific Research on Innovative
Areas No.\ 21111006 (JY), the Wallonia-Brussels Federation grant ARC
11/15-040 and the ESA Belgian Federal PRODEX program 
$\mathrm{N}^\circ 4000103071$ (CR).

\bibliographystyle{JHEP}
\bibliography{magneticfields,reheating}

\end{document}